# Low-latency adiabatic quantum-flux-parametron circuit integrated with a hybrid serializer/deserializer

Yuki Hironaka[1], Taiki Yamae[1,2], Christopher L. Ayala[3], (Senior Member, IEEE), Nobuyuki Yoshikawa[1,3], (Senior Member, IEEE), and Naoki Takeuchi[4]

[1]Department of Electrical and Computer Engineering, Yokohama National University, Yokohama, Kanagawa 240-8501, Japan
[2]Research Fellow of Japan Society for the Promotion of Science, Chiyoda, Tokyo 102-0083, Japan
[3]Institute of Advanced Sciences, Yokohama National University, Yokohama, Kanagawa 240-8501, Japan
[4]Research Center for Emerging Computing Technologies, National Institute of Advanced Industrial Science and Technology (AIST), Tsukuba, Ibaraki 305-8560 Japan

Corresponding author: Naoki Takeuchi (e-mail: n-takeuchi@aist.go.jp).

This work was supported by JSPS KAKENHI (Grant Numbers JP22H00220, JP19H05614, and JP18H05245).

**ABSTRACT** Adiabatic quantum-flux-parametron (AQFP) logic is an ultra-low-power superconductor logic family. AQFP logic gates are powered and clocked by dedicated clocking schemes using ac excitation currents to implement an energy-efficient switching process, adiabatic switching. We have proposed a low-latency clocking scheme, delay-line clocking, and demonstrated basic AQFP logic gates. In order to test more complex circuits, a serializer/deserializer (SerDes) should be incorporated into the AQFP circuit under test, since the number of input/output (I/O) cables is limited by equipment. Therefore, in the present study we propose and develop a novel SerDes for testing delay-line-clocked AQFP circuits by combining AQFP and rapid single-flux-quantum (RSFQ) logic families, which we refer to as the AQFP/RSFQ hybrid SerDes. The hybrid SerDes comprises RSFQ shift registers to facilitate the data storage during serial-to-parallel and parallel-to-serial conversion. Furthermore, all the component circuits in the hybrid SerDes are clocked by the identical excitation current to synchronize the AQFP and RSFQ parts. We fabricate and demonstrate a delay-line-clocked AQFP circuit (8-to-3 encoder, which is the largest delay-line-clocked circuit ever designed) integrated with the hybrid SerDes at 4.2 K up to 4.5 GHz. Our measurement results indicate that the hybrid SerDes enables the testing of delay-line-clocked AQFP circuits with only a few I/O cables and is thus a powerful tool for the development of very large-scale integration AQFP circuits.

## I. INTRODUCTION

Superconductor Josephson logic families, such as rapid single-flux-quantum (RSFQ) logic [1] and its energy-efficient variants [2, 3], can operate with much smaller energy dissipation than that of conventional semiconductor circuits and are thus promising technology for applications requiring high energy efficiency: low-power microprocessors [4–6], readout circuits for superconducting detectors [7–10], and interface circuits for superconducting qubits [11, 12]. Among various Josephson logic families, adiabatic quantum-flux-parametron (AQFP) logic [13, 14] exhibits extremely high energy-efficiency. Due to adiabatic switching [15–17], AQFP circuits can operate with a small energy dissipation of approximately $10^{-21}$ J per junction at a 5 GHz clock frequency [18], which is much smaller than that of conventional Josephson logic families ($\sim 10^{-19}$ J). We have developed and demonstrated various AQFP circuits to show the potential to be widely used for future information and communication technology [14].

AQFP circuits are operated by multi-phase clocking because all AQFP logic gates must be clocked in the order of logical operations by ac excitation currents. Typically, we utilize four-phase clocking [19, 20] to demonstrate AQFP circuits, where AQFP logic gates are clocked with a phase separation of 90° by paired excitation currents. The drawback of four-phase clocking is the relatively long latency, which is determined by the phase difference of clocking (i.e., a quarter clock cycle) rather than the switching speed of each logic gate. Assuming a 5-GHz clock frequency, the latency of an AQFP logic gate is 50 ps, which is rather long compared to that of other Josephson logic families. Moreover, the latency of an AQFP 8-bit carry look-ahead adder using four-phase clocking was reported to be 800 ps at 5 GHz [18], which is much longer than that of its counterpart using reciprocal quantum logic (150 ps) [21]. Therefore, to reduce the latency of AQFP circuits, we proposed a low-latency clocking scheme, delay-line clocking [22], where AQFP logic gates are clocked by a single ac excitation current, and the latency of each logic gate is determined by the propagation delay of the excitation current. The latency of delay-line clocking can be set to much shorter values than that of four-phase clocking. We demonstrated basic AQFP logic gates using delay-line clocking with a latency as short as 10–20 ps per gate [22, 23].

In the present study, we propose and demonstrate a serializer/deserializer (SerDes) toward the testing of large-scale AQFP circuits with delay-line clocking. A SerDes, a pair of a serializer (parallel-to-serial converter) and a deserializer (serial-to-parallel converter), is an important circuit block in cryogenic experiments. For instance, the number of available input/output (I/O) cables is limited by equipment such as a cryostat and cryoprobe [24], so that it is crucial to reduce the number of I/O cables as much as possible using a Ser/Des, especially when testing a large-scale superconductor circuit. In RSFQ logic, serializers and deserializers are implemented by shift registers [25, 26], which store data during serial-to-parallel (S2P) and parallel-to-serial (P2S) conversion. As for AQFP logic, we previously proposed and demonstrated the feedback-type SerDes [27], where an AQFP buffer chain with feedback paths operates in a similar way to a shift register. However, this Ser/Des was developed for four-phase clocking and does not operate in delay-line clocking, because feedback paths are difficult to make in delay-line clocking due to the low latency. Therefore, we develop a novel SerDes for testing delay-line-clocked AQFP circuits by combining AQFP and RSFQ technologies, which we refer to as the AQFP/RSFQ hybrid SerDes.

First, we briefly describe the conventional SerDes for four-phase-clocked AQFP circuits, i.e., feedback-type SerDes. Then, we explain the details of the hybrid SerDes with a focus on the following two points: (i) The hybrid SerDes comprises RSFQ shift registers to facilitate the data storage during S2P and P2S conversion, with AQFP/RSFQ interfaces [28, 29] transmitting data between the shift registers and the AQFP circuit under test (CUT). (ii) All the component circuits are seamlessly clocked by the identical excitation current, which facilitates the synchronization between the RSFQ and AQFP parts. Finally, we demonstrate a delay-line-clocked AQFP circuit (8-to-3 encoder) integrated with the hybrid SerDes at 4.2 K up to 4.5 GHz in order to validate that the hybrid SerDes enables the testing of delay-line-clocked AQFP circuits with only a few I/O cables and is thus a powerful tool toward the development of very large-scale integration AQFP circuits.

## II. CONVENTIONAL DESIGN: FEEDBACK-TYPE SERDES

Figure 1(a) depicts a circuit diagram of the feedback-type deserializer [27], converting the serial input ($D$) into the 4-bit parallel outputs ($Q_0$ through $Q_3$). This deserializer comprises basic AQFP logic gates, clocked by paired ac excitation currents ($I_q$ and $I_i$) in the manner of four-phase clocking [20] (the dc offset current is omitted for simplicity). $\phi_1$ through $\phi_4$ denote the excitation phases, along which data transmit with a phase separation of 90°. The deserializer includes four bit slices, each with seven buffers and an AND gate. The bit slices are connected in series by the feedback paths from $\phi_4$ to $\phi_1$, thus operating as a shift register with each bit slice storing 1-bit data. The serial input transmits through the bit slices from $D$ in synchronization with $I_q$ and $I_i$. After the serial input is loaded into the bit slices, a logic 1 is applied to the enable-signal port ($E$), so that the serial input from $D$ is converted into the parallel outputs at $Q_0$ through $Q_3$. Note that $Q_0$ through $Q_3$ are always zeros when $E$ is a zero, due to the AND gate in each bit slice.

Figure 1(b) depicts a circuit diagram of the feedback-type serializer [27], converting the 4-bit parallel inputs ($D_0$ through $D_3$) into the serial output ($Q$). As with the deserializer, this serializer comprises basic AQFP logic gates and is clocked by $I_q$ and $I_i$. Between each pair of adjacent inputs, a buffer chain with a feedback path of an appropriate length is inserted to delay data. Consequently, the parallel inputs from $D_0$ through $D_3$ are converted into the serial output at $Q$ when a logic 1 is applied to the enable-signal port ($E$).

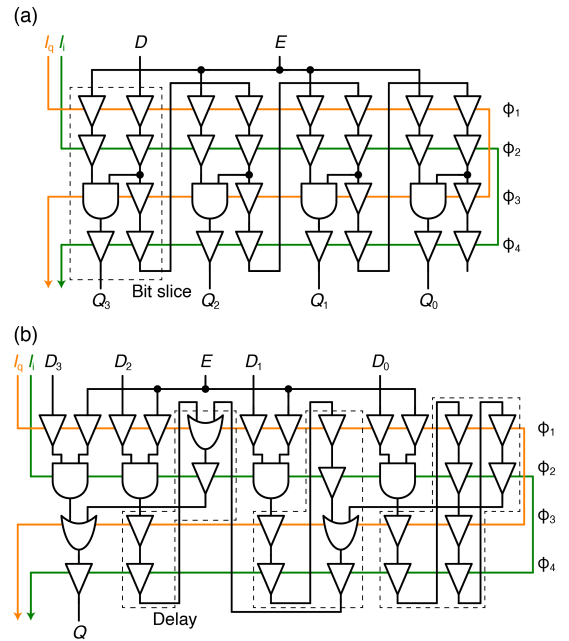

**FIGURE 1.** Feedback-type SerDes for four-phase clocking. (a) Deserializer converting the serial input ($D$) into 4-bit parallel outputs ($Q_0$ through $Q_3$). (b) Serializer converting the 4-bit parallel inputs ($D_0$ through $D_3$) into serial output ($Q$).

As shown above, the feedback-type SerDes utilizes feedback paths from $\phi_4$ to $\phi_1$ in order to store data during S2P and P2S conversion. In typical AQFP circuits, feedback paths are made between the final and initial excitation phases, so that feedback can be easily made in clocking with a small number of excitation phases, such as four-phase clocking. On the other hand, delay-line clocking operates like many-phase clocking due to the low latency and is not suitable for implementing the feedback-type SerDes; thus, a different type of SerDes is required for performing S2P and P2S conversion in delay-line clocking.

## III. AQFP/RSFQ HYBRID SERDES

### A. DESERIALIZER

The AQFP/RSFQ hybrid SerDes utilizes RSFQ shift registers to store data during S2P and P2S conversion. Figure 2(a) depicts a block diagram of the hybrid deserializer, comprising AQFP-to-RSFQ (A2R) and RSFQ-to-AQFP (R2A) interfaces [28], [29] and basic AQFP logic gates (const-1 gates [30] and buffers). The non-shaded and shaded regions represent AQFP and RSFQ parts, respectively. The R2A interfaces form a shift register because an R2A interface includes an RSFQ D flip-flop (DFF). The A2R interfaces and const-1 gates generate single-flux-quantum (SFQ) clock signals for operating the R2A-based shift register. To facilitate the synchronization between the AQFP and RSFQ parts, all the component circuits are clocked by a single excitation current $I_x$ in the manner of delay-line clocking [22], where logic operations are performed with latency determined by the propagation delay of the delay-line inserted between each pair of adjacent excitation phases. The operation of the hybrid deserializer is as follows: The const-1 gates generate logic-1 signals and apply them to the A2R interfaces, and the A2R interfaces convert the logic-1 signals into SFQ pulses for clocking the R2A-based shift register, thereby shifting the data in the shift register. The serial input $D$ is applied to the first R2A interface in the shift register in synchronization with $I_x$. After the serial input is loaded into the R2A interfaces, the enable-signal current $I_{ena}$ is serially applied to the R2A interfaces. Consequently, the AQFP buffers receive the signals from the R2A interfaces and generate 4-bit parallel outputs, $Q_0$ through $Q_3$. Figure 2(b) shows simulation waveforms of the hybrid deserializer with 20-ps delay lines at 5 GHz, where $V_D$ is the input voltage applied to $D$ and $I_{Q0}$ through $I_{Q3}$ are the signal currents representing $Q_0$ through $Q_3$, respectively. The dashed lines represent a zero for each waveform. The simulation was conducted using a Josephson circuits simulator, JSIM [31], and the process-specific parameters for the AIST

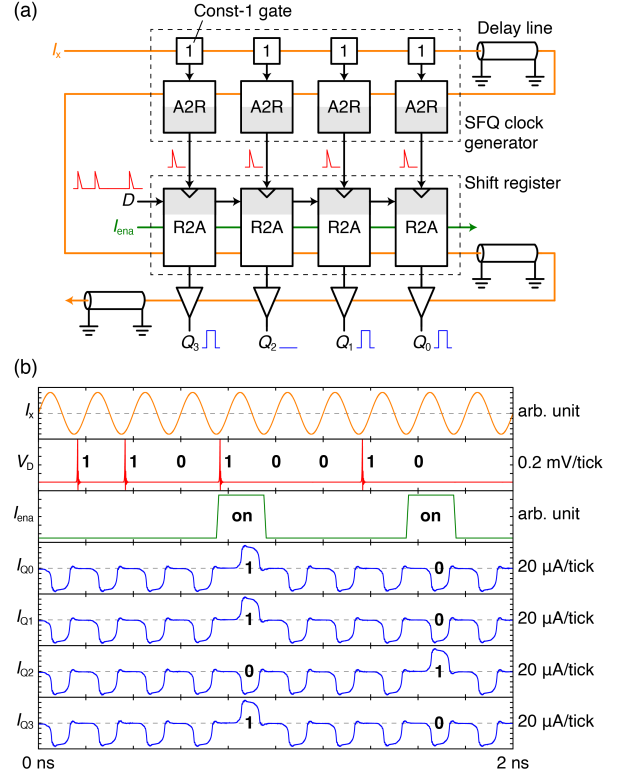

**FIGURE 2.** Hybrid deserializer for delay-line clocking. (a) Block diagram and (b) simulation waveforms at 5 GHz. The serial input ($V_D$) is converted into the 4-bit parallel outputs ($I_{Q0}$ through $I_{Q3}$) in synchronization with $I_x$.

10 kA/cm$^2$ Nb high-speed standard process (HSTP) [20]. The simulation waveforms show that two serial inputs of "1101" and "0010" are converted into the corresponding parallel outputs.

Here we give more explanations for the A2R-based SFQ clock generator and R2A-based shift register. $I_x$ applies an ac excitation flux with an amplitude of $0.5\Phi_0$ to each AQFP gate, where $\Phi_0$ is the flux quantum, and a dc offset current $I_d$ [not shown in Fig. 2(a)] applies a dc offset flux of $0.5\Phi_0$ to each AQFP gate. The clock generator is operated with a negative $I_d$ whereas the other circuits are operated with a positive $I_d$. As a result, the clock generator and shift register are clocked by $I_x$ at different timings: the clock generator is clocked at the fall edge of $I_x$ whereas the shift register is clocked at the rise edge of $I_x$ (i.e., half clock cycle later). This is the reason why a delay line is not inserted between the A2R and R2A interfaces. Moreover, the clock generator produces parallel SFQ clocks with small skews along with $I_x$, compared to the typical SFQ clock distribution via a splitter network [32]. The above features ensure that the shift register generates outputs after all data are shifted.

Next, we describe the details of the A2R and R2A interfaces. The design of the A2R interfaces is the same as that of the compact A2R interface shown in the literature [29], which can save the junction count and footprint compared to the original design [28]. The design of the R2A interfaces is based on that shown in the literature [28] but was slightly modified such that $I_{ena}$ can activate the R2A interface. Figure 3(a) depicts a simplified circuit diagram of the R2A interface used in the present study. The R2A interface is composed of an RSFQ DFF and AQFP buffer, where the storage loop in the DFF is coupled to the input branch of the buffer. $I_{ena}$ applies a negative offset flux to the input branch such that the buffer switches to a logic 1 only when an SFQ is kept in the storage loop and $I_{ena}$ is on. Figure 3(b) shows simulation waveforms of the R2A interface at 5 GHz, where $V_{in}$ ($V_{clk}$) is the SFQ input (clock) applied to the RSFQ DFF and $I_s$ is the signal current of the AQFP buffer. This figure shows that $V_{in}$ is converted into $I_s$ when $I_{ena}$ is on, and that $I_s$ always represents a logic 0 when $I_{ena}$ is off. In the present design, the value of $I_{ena}$ is $-130$ µA and $-390$ µA when $I_{ena}$ is on and off, respectively.

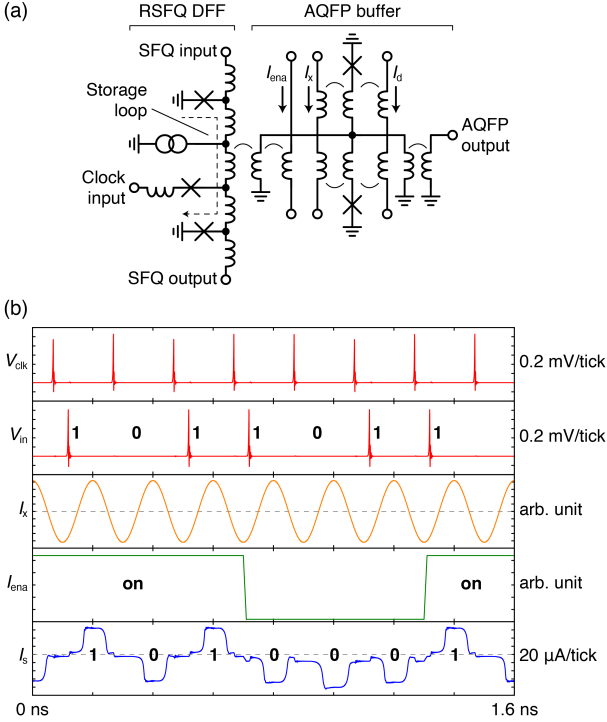

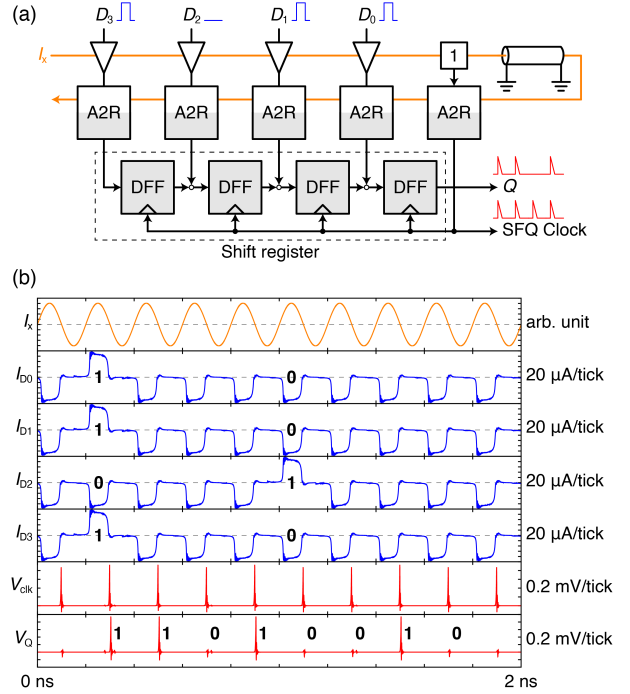

FIGURE 3. RSFQ-to-AQFP (R2A) interface. (a) Circuit diagram and (b) simulation waveforms at 5 GHz. When the enable signal ($I_{ena}$) is on, the SFQ input ($V_{in}$) is converted into the signal current ($I_s$) in an AQFP buffer.

FIGURE 4. Hybrid serializer for delay-line clocking. (a) Block diagram and (b) simulation waveforms at 5 GHz. The parallel inputs ($I_{D0}$ through $I_{D3}$) are converted into the serial output ($V_Q$) in synchronization with $I_x$.

### B. SERIALIZER

Figure 4(a) depicts a block diagram of the hybrid serializer, comprising A2R interfaces, RSFQ DFFs, and basic AQFP logic gates (a const-1 gate and buffers). The AQFP buffers receive the parallel inputs, $D_0$ through $D_3$, and the A2R interfaces convert $D_0$ through $D_3$ into the corresponding SFQ pulses, which are loaded into the DFF-based shift register. The const-1 gate and an A2R interface generate SFQ clock signals for operating the DFFs in synchronization with $I_x$. As a consequence, $D_0$ through $D_3$ are converted into the serial output $Q$ from the shift register. In experiments, $Q$ is amplified by a voltage driver using a stack of dc superconducting quantum interference devices (SQUIDs) [33]; thus, the SFQ clock signals are also output to operate this voltage driver. Moreover, since we assume that the serializer is used in combination with the deserializer, an enable signal is not needed to control the serializer; the enable signal applied to the deserializer blocks unwanted S2P and P2S conversion at the very start. Figure 4(b) shows simulation waveforms of the hybrid serializer with a 20-ps delay line at 5 GHz, where $I_{D0}$ through $I_{D3}$ are the signal currents representing $D_0$ through $D_3$, respectively, $V_{clk}$ is the SFQ clock produced by an A2R interface, and $V_Q$ is the output voltage representing $Q$. The waveforms show that the two parallel inputs of "1101" and "0010" are converted into the corresponding serial outputs.

### IV. EXPERIMENTS

To validate the functionality of the AQFP/RSFQ hybrid SerDes, we fabricated and demonstrated a delay-line-clocked AQFP circuit (8-to-3 encoder) integrated with the hybrid SerDes. Figure 5(a) shows a micrograph of the encoder chip fabricated by the HSTP. The operation of this chip is as follows: The input current $I_{in}$ applies an 8-bit one-hot serial input to the deserializer, and the deserializer converts the serial input into the corresponding parallel outputs. The encoder receives the 8-bit one-hot parallel

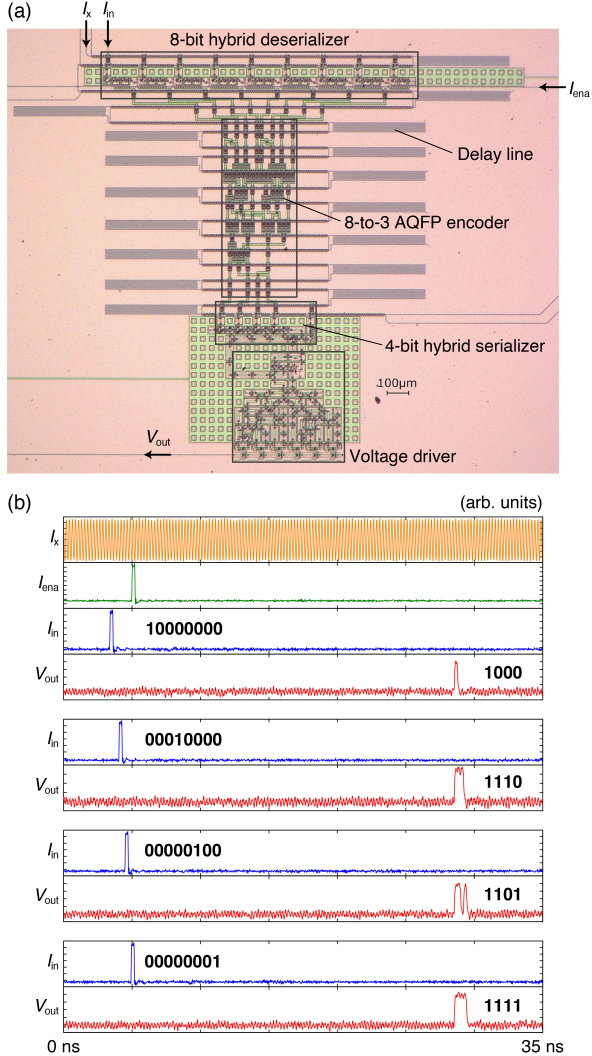

**FIGURE 5.** AQFP encoder integrated with a hybrid SerDes. (a) Micrograph of a chip and (b) measurement waveforms at 4.5 GHz for four different input patterns.

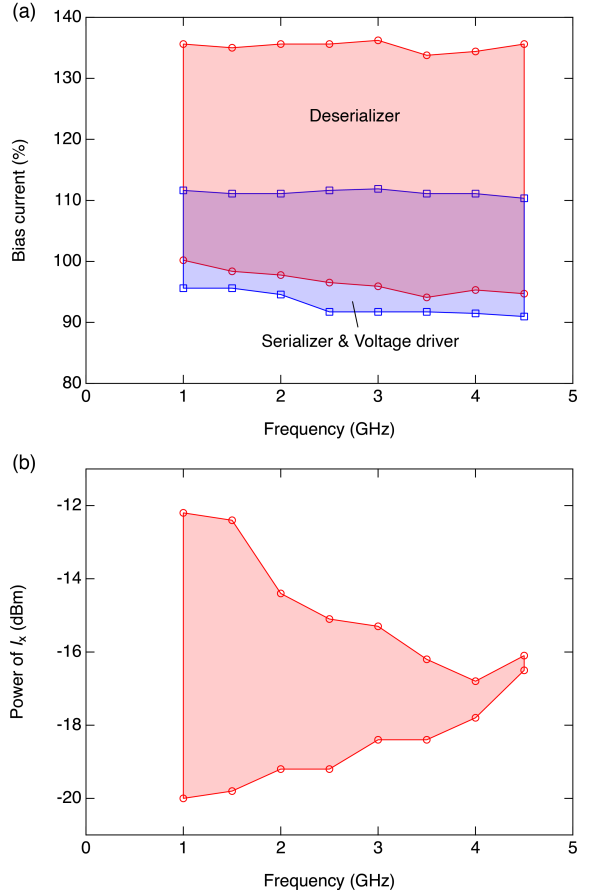

**FIGURE 6.** Measurement results of the operating margins regarding (a) the dc bias currents applied to the deserializer, serializer, and voltage driver and (b) the ac excitation current.

inputs from the deserializer and generates 3-bit outputs when $I_{ena}$ is on. The encoder also produces a flag bit that tells the beginning of the data sequence. Then, the serializer receives the 4-bit parallel inputs from the encoder and generates a 4-bit serial output, which is amplified into mV-range voltage signals ($V_{out}$) by the SQUID-stack voltage driver [33] and is observed at the room-temperature stage. Consequently, the encoder can be tested using only a few I/O cables. The entire circuit was designed using AQFP and RSFQ cell libraries developed for the HSTP [20, 34] and includes 742 Josephson junctions: 207 for the deserializer, 200 for the encoder, 113 for the serializer, and 222 for the voltage driver. Moreover, a delay line with a latency of approximately 20 ps was inserted into each pair of adjacent excitation phases, except between the A2R and R2A interfaces in the deserializer [see Fig. 2(a)]. The encoder includes 13 excitation phases, and thus its latency is 20 ps ×

13 = 260 ps. The power dissipation of the encoder is estimated to be 621 pW at 4.5 GHz (the highest operating frequency in the experiments) by JSIM.

We tested the encoder chip using a wideband cryoprobe [24] at 4.2 K in liquid He. During experiments, $I_x$ was provided by a vector signal generator (Anritsu, MG3710A), and $I_{in}$ and $I_{ena}$ were provided by a pulse pattern generator (Anritsu, MU181020B), where the signal generator also produced clock signals to synchronize with the pattern generator. Figure 5(b) shows measurement waveforms of the encoder chip at 4.5 GHz for four different input patterns, observed by a wideband oscilloscope (Keysight, DSOV164A). For example, the second pair of $I_{in}$ and $V_{out}$ show that a one-hot input of "00010000" is converted into an output of "1110," where the first bit is a flag bit and the other three bits (110, with the least significant bit appearing first) indicate the location of a logic 1 in the input. We confirmed that the encoder chip operates for all the eight one-hot input patterns (10000000, 01000000, ... , 00000001) up to 4.5 GHz, thereby validating the functionality of the hybrid SerDes. Note that the long latency between $I_{ena}$ and $V_{out}$ shown in Fig. 5(b) is due to the propagation delay through equipment such as the cryoprobe, rather than the latency of the encoder chip.

Figure 6(a) shows the measured operating margins of the dc bias currents applied to the deserializer, serializer, and voltage driver as functions of the operating frequency, where the serializer and voltage driver were powered by the common bias current. All the eight input patterns were tested at each operating frequency. The shaded regions represent where the circuits operate, with the bias currents normalized by the design values (17.8 mA for the deserializer and 38.6 mA for the serializer and voltage driver). Both operating margins are reasonably wide for the entire frequency range but shift to the upper side. This may be due to the characteristics of the A2R interfaces [29], which require relatively high bias currents to transfer SFQ pulses generated by a non-adiabatic quantum-flux-parametron gate into an RSFQ gate. Figure 6(b) shows the measured operating margins of $I_x$, which represent the noise margins of the encoder. The operating margin is reasonably wide up to 3 GHz but shrinks significantly at higher frequencies in a similar way to the previous experiment [23]. This may be due to process variation and indicates the necessity of more robust circuit design for even higher operating frequencies.

## V. CONCLUSIONS

We proposed the AQFP/RSFQ hybrid SerDes toward the testing of large-scale AQFP circuits with delay-line clocking. The hybrid SerDes comprises RSFQ shift registers for storing data during S2P and P2S conversion and AQFP/RSFQ interfaces for transmitting data between the shift registers and the AQFP CUT. Moreover, all the component circuits in the hybrid SerDes are seamlessly clocked by a single excitation current to synchronize the AQFP and RSFQ parts. We fabricated and tested an 8-to-3 encoder integrated with the hybrid SerDes, which is the largest delay-line-clocked AQFP circuit ever designed, at 4.2 K up to 4.5 GHz, thereby demonstrating that the hybrid SerDes enables the testing of delay-line-clocked AQFP circuits with only a few I/O cables at high clock frequencies. Our next step is to demonstrate even larger AQFP circuits using the hybrid SerDes.

## ACKNOWLEDGMENT

The circuits were fabricated in the clean room for analog-digital superconductivity (CRAVITY) of National Institute of Advanced Industrial Science and Technology (AIST) with the high-speed standard process (HSTP). We would like to thank S. Miyajima and F. China for providing the SQUID-stack voltage driver and C. J. Fourie for providing a 3D inductance extractor, InductEx.

## REFERENCES

[1] K. K. Likharev and V. K. Semenov, "RSFQ logic/memory family: a new Josephson-junction technology for sub-terahertz-clock-frequency digital systems," *IEEE Trans. Appl. Supercond.*, vol. 1, no. 1, pp. 3–28, Mar. 1991.
[2] D. E. Kirichenko, S. Sarwana, and A. F. Kirichenko, "Zero static power dissipation biasing of RSFQ circuits," *IEEE Trans. Appl. Supercond.*, vol. 21, no. 3, pp. 776–779, Jun. 2011.
[3] M. Tanaka, A. Kitayama, T. Koketsu, M. Ito, and A. Fujimaki, "Low-energy consumption RSFQ circuits driven by low voltages," *IEEE Trans. Appl. Supercond.*, vol. 23, no. 3, p. 1701104, Jun. 2013.
[4] A. F. Kirichenko et al., "ERSFQ 8-bit parallel arithmetic logic unit," *IEEE Trans. Appl. Supercond.*, vol. 29, no. 5, p. 1302407, Aug. 2019.
[5] R. Sato et al., "High-speed operation of random-access-memory-embedded microprocessor with minimal instruction set architecture based on rapid single-flux-quantum logic," *IEEE Trans. Appl. Supercond.*, vol. 27, no. 4, p. 1300505, Jun. 2017.
[6] C. L. Ayala, T. Tanaka, R. Saito, M. Nozoe, N. Takeuchi, and N. Yoshikawa, "MANA: a Monolithic Adiabatic iNtegration Architecture microprocessor using 1.4-zJ/op unshunted superconductor Josephson junction devices," *IEEE J. Solid-State Circuits*, vol. 56, no. 4, pp. 1152–1165, Apr. 2021.
[7] H. Terai, S. Miki, and Z. Wang, "Readout electronics using single-flux-quantum circuit technology for superconducting single-photon detector array," *IEEE Trans. Appl. Supercond.*, vol. 19, no. 3, pp. 350–353, Jun. 2009.
[8] A. Sahu, M. E. Celik, D. E. Kirichenko, T. V Filippov, and D. Gupta, "Low-power digital readout circuit for superconductor nanowire single-photon detectors," *IEEE Trans. Appl. Supercond.*, vol. 29, no. 5, p. 1301306, Aug. 2019.
[9] N. Takeuchi et al., "Scalable readout interface for superconducting nanowire single-photon detectors using AQFP and RSFQ logic families," *Opt. Express*, vol. 28, no. 11, pp. 15824–15834, May 2020.
[10] M. Tanaka et al., "SFQ parallel encoders promising for video imaging with superconductor stripline detectors," *IEEE Trans. Appl. Supercond.*, vol. 31, no. 1, p. 1300106, Jan. 2021.
[11] E. Leonard et al., "Digital coherent control of a superconducting qubit," *Phys. Rev. Appl.*, vol. 11, no. 1, p. 014009, 2019.
[12] L. Howe et al., "Digital control of a superconducting qubit using a Josephson pulse generator at 3 K," *PRX Quantum*, vol. 3, no. 1, p. 010350, Mar. 2022.
[13] N. Takeuchi, D. Ozawa, Y. Yamanashi, and N. Yoshikawa, "An adiabatic quantum flux parametron as an ultra-low-power logic device," *Supercond. Sci. Technol.*, vol. 26, no. 3, p. 035010, Mar. 2013.
[14] N. Takeuchi, T. Yamae, C. L. Ayala, H. Suzuki, and N. Yoshikawa, "Adiabatic quantum-flux-parametron: A tutorial review," *IEICE Trans. Electron.*, vol. E105.C, no. 6, p. 2021SEP0003, Jun. 2022.
[15] R. W. Keyes and R. Landauer, "Minimal energy dissipation in logic," *IBM J. Res. Dev.*, vol. 14, no. 2, pp. 152–157, Mar. 1970.
[16] K. K. Likharev, "Classical and quantum limitations on energy consumption in computation," *Int. J. Theor. Phys.*, vol. 21, no. 3–4, pp. 311–326, Apr. 1982.
[17] J. G. Koller and W. C. Athas, "Adiabatic switching, low energy computing, and the physics of storing and erasing information," in *Workshop on Physics and Computation*, 1992, pp. 267–270.
[18] N. Takeuchi, T. Yamae, C. L. Ayala, H. Suzuki, and N. Yoshikawa, "An adiabatic superconductor 8-bit adder with $24k_BT$ energy dissipation per junction," *Appl. Phys. Lett.*, vol. 114, no. 4, p. 042602, Jan. 2019.
[19] W. Hioe, M. Hosoya, S. Kominami, H. Yamada, R. Mita, and K. Takagi, "Design and operation of a quantum flux parametron bit-slice ALU," *IEEE Trans. Appl. Supercond.*, vol. 5, no. 2, pp. 2992–2995, Jun. 1995.
[20] N. Takeuchi et al., "Adiabatic quantum-flux-parametron cell library designed using a 10 kA cm$^{-2}$ niobium fabrication process," Supercond. Sci. Technol., vol. 30, no. 3, p. 035002, Mar. 2017.
[21] A. Y. Herr et al., "An 8-bit carry look-ahead adder with 150 ps latency and sub-microwatt power dissipation at 10 GHz," *J. Appl. Phys.*, vol. 113, no. 3, p. 033911, 2013.

[22] N. Takeuchi, M. Nozoe, Y. He, and N. Yoshikawa, "Low-latency adiabatic superconductor logic using delay-line clocking," *Appl. Phys. Lett.*, vol. 115, no. 7, p. 072601, Aug. 2019.

[23] T. Yamae, N. Takeuchi, and N. Yoshikawa, "Adiabatic quantum-flux-parametron with delay-line clocking: Logic gate demonstration and phase skipping operation," *Supercond. Sci. Technol.*, vol. 34, no. 12, p. 125002, Oct. 2021.

[24] H. Suzuki, N. Takeuchi, and N. Yoshikawa, "Development of the wideband cryoprobe for evaluating superconducting integrated circuits," *IEICE Trans. Electron. (Japanese Ed.)*, vol. J104-C, no. 6, pp. 193–201, 2021.

[25] H. Park, Y. Yamanashi, K. Taketomi, N. Yoshikawa, A. Fujimaki, and N. Takagi, "Novel serial–parallel converter using SFQ logic circuits," *Phys. C Supercond.*, vol. 468, no. 15–20, pp. 1977–1982, Sep. 2008.

[26] S. Miyajima, M. Yabuno, S. Miki, T. Yamashita, and H. Terai, "High-time-resolved 64-channel single-flux quantum-based address encoder integrated with a multi-pixel superconducting nanowire single-photon detector," *Opt. Express*, vol. 26, no. 22, p. 29045, Oct. 2018.

[27] C. L. Ayala, N. Takeuchi, and N. Yoshikawa, "Adiabatic Quantum-Flux-Parametron Design-For-Testability Components for Large-Scale Digital Circuits," in *32nd International Symposium on Superconductivity (ISS2019)*, EDP2-7, Kyoto.

[28] F. China et al., "Demonstration of signal transmission between adiabatic quantum-flux-parametrons and rapid single-flux-quantum circuits using superconductive microstrip lines," *IEEE Trans. Appl. Supercond.*, vol. 27, no. 4, p. 1300205, 2017.

[29] Y. Yamazaki, N. Takeuchi, and N. Yoshikawa, "A compact interface between adiabatic quantum-flux-parametron and rapid single-flux-quantum circuits," *IEEE Trans. Appl. Supercond.*, vol. 31, no. 5, p. 1302705, Aug. 2021.

[30] T. Ando, N. Takeuchi, Y. Yamanashi, and N. Yoshikawa, "Adiabatic quantum-flux-parametron constant cells using asymmetrical structures," *IEEJ Trans. Fundam. Mater.*, vol. 136, no. 12, pp. 747–752, 2016. (in Japanese)

[31] E. Fang and T. Van Duzer, "A Josephson Integrated Circuit Simulator (JSIM) for Superconductive Electronics Application," in *1989 International Superconductivity Electronics Conference (ISEC '89)*, 1989, pp. 407–410.

[32] M. Tanaka, R. Sato, Y. Hatanaka, and A. Fujimaki, "High-density shift-register-based rapid single-flux-quantum memory system for bit-serial microprocessors," *IEEE Trans. Appl. Supercond.*, vol. 26, no. 5, p. 1301005, Aug. 2016.

[33] Y. Hashimoto, H. Suzuki, S. Nagasawa, M. Maruyama, K. Fujiwara, and M. Hidaka, "Measurement of superconductive voltage drivers up to 25 Gb/s/ch," *IEEE Trans. Appl. Supercond.*, vol. 19, no. 3, pp. 1022–1025, Jun. 2009.

[34] N. Takeuchi, H. Suzuki, C. J. Fourie, and N. Yoshikawa, "Impedance design of excitation lines in adiabatic quantum-flux-parametron logic using InductEx," *IEEE Trans. Appl. Supercond.*, vol. 31, no. 5, p. 1300605, Aug. 2021.